\documentclass[conference]{IEEEtran}
\IEEEoverridecommandlockouts
\usepackage{cite}
\usepackage{amsmath,amssymb,amsfonts}
\usepackage{algorithmic}
\usepackage{graphicx}
\usepackage{textcomp}
\usepackage{xcolor}
\usepackage{url} % For URLs in references
\usepackage{listings}
\usepackage{booktabs}   % 用于 \toprule, \midrule, \bottomrule 等命令
\usepackage{multirow}   % 用于处理多行单元格 (虽然此表没用，但最好有)
\usepackage{tabularx}   % 这是解决表格换行问题的关键宏包

\def\BibTeX{{\rm B\kern-.05em{\sc i\kern-.025em b}\kern-.08em
    T\kern-.1667em\lower.7ex\hbox{E}\kern-.125emX}}
\begin{document}

\title{White-Box Reasoning: Synergizing LLM Strategy and gm/Id Data for Automated Analog Circuit Design}

\author{
    \IEEEauthorblockN{
        1\textsuperscript{st} Jianqiu Chen\IEEEauthorrefmark{1},
        2\textsuperscript{nd} Siqi Li\IEEEauthorrefmark{1}, and
        3\textsuperscript{rd} Xu He\IEEEauthorrefmark{2}
    }
    \IEEEauthorblockA{
        \IEEEauthorrefmark{1}Shanghai Hynitron Technology Co., Ltd, Shanghai, China \\
        Email: jianqiu.chen@hynitron.com.cn
    }
    \IEEEauthorblockA{
        \IEEEauthorrefmark{2}College of Information Science and Engineering, Hunan University, Changsha, China
    }
}

\maketitle

\begin{abstract}
Analog integrated circuit design has long been challenged by its reliance on engineers' personal experience and inefficient iterative SPICE simulations, creating a significant bottleneck in automation. The root cause is the inability of traditional theoretical formulas to accurately guide design practices in advanced process nodes. Although Large Language Models (LLMs) have shown immense potential as powerful reasoning engines, their application risks devolving into a more sophisticated form of "guessing" without a foundation in rigorous engineering principles. To address this challenge, this paper presents an innovative "synergistic reasoning" framework that, for the first time, deeply integrates the macro-level strategic reasoning of an LLM with the micro-level physical precision of the industry-standard gm/Id design methodology. In this framework, the LLM is empowered with pre-computed gm/Id lookup tables, transforming it from a qualitative heuristic advisor into a quantitative, data-driven design partner.

As a validation, we apply this framework to the design of a classic two-stage operational amplifier. Experiments show that under our framework, the Gemini model successfully converges and meets all specifications at the TT corner in just 5 iterations. We further demonstrate the framework's capability to extend this optimization to cover all PVT corners. Crucially, a pivotal ablation study proves that the injection of gm/Id data is the key to achieving efficient and precise reasoning; without it, the LLM's optimization process is significantly slower and more prone to deviation. Furthermore, when compared to a design by a senior engineer, our framework achieves a "quasi-expert" level of design quality while delivering an order-of-magnitude improvement in efficiency. This work validates a viable path for combining the powerful reasoning of LLMs with scientific circuit design methodologies, offering a novel and feasible approach to achieving true analog design automation.
\end{abstract}


\section{Introduction}

\subsection{The Automation Dilemma in Analog Circuit Design}
Unlike digital integrated circuit design, which has achieved a high degree of automation, analog design workflows remain labor-intensive and heavily dependent on engineers' personal experience. This "automation gap" has become a critical bottleneck limiting the innovation efficiency of entire electronic systems. To address this challenge, academia and industry have proposed various automated design methods over the past few decades, such as tools based on Genetic Algorithms (GA)~\cite{dastidar2010automated}, Particle Swarm Optimization (PSO), Bayesian Optimization (BO)~\cite{hakhamaneshi2019efficient}, and Reinforcement Learning (RL).

However, these traditional automation algorithms often face challenges of slow convergence and low efficiency in practical applications. The fundamental reason is that they mostly treat the complex analog circuit as a "black box". These algorithms search within a high-dimensional, highly-constrained parameter space to find an optimal solution that meets performance specifications, essentially reducing the task to a pure mathematical optimization problem. This "black box" approach ignores the deep physical principles and design heuristics within the circuit, causing the search process to be "akin to finding a needle in a haystack," often requiring hundreds or thousands of simulation iterations to converge, which is prohibitively time-consuming.

\subsection{Breaking the "Black Box": The Injection of Domain Knowledge}
We argue that to achieve truly efficient automation, the primary task must be to break the "black box" by injecting domain knowledge into the optimization loop. For modern analog circuit design, especially in advanced process nodes, traditional textbook formulas are no longer accurate. Therefore, the most reliable domain knowledge comes from an accurate characterization of the process's physical properties.

The well-established \textit{gm/Id} design methodology~\cite{silveira1996gm} provides an effective approach to address this challenge. By using pre-computed lookup tables generated from simulations, it accurately describes a transistor's performance parameters, replacing inaccurate analytical formulas with precise physical data. Injecting \textit{gm/Id} data into the automation flow means the optimizer is no longer probing blindly but navigating based on physical reality, which can significantly prune the search space and improve efficiency.

\subsection{The Strategy Challenge: From "Data" to "Decision"}
Even with a precise "physics-based handbook" provided by the gm/Id method, the path to automation still faces a final, crucial challenge. Analog circuit design is not a simple table-lookup calculation but a complex decision-making process involving dynamic trade-offs among multiple objectives (e.g., gain, bandwidth, power, noise). When multiple performance metrics conflict, which one should be prioritized for optimization? How should parameters be adjusted to achieve the best overall effect?

This requires a "brain" capable of flexible, multi-objective trade-offs, much like a senior engineer. Traditional, rule-based scripts or algorithms are ill-suited for such highly dynamic and heuristic strategic tasks. This, however, is precisely where Large Language Models (LLMs) excel. The academic community has begun to explore LLM applications in circuit design, achieving preliminary results in areas such as topology generation~\cite{liu2023analogxpert}, code generation~\cite{xu2024analogcoder}, and integration with traditional optimization algorithms~\cite{yin2024ado}. With their powerful contextual understanding and complex reasoning abilities, LLMs have become the ideal candidates to play the role of this "strategy engine."

\subsection{Our Contribution: A Synergistic Solution of "Brain" and "Handbook"}
Based on the analysis above, this paper proposes and implements an innovative "synergistic reasoning" framework. We do not use the LLM as a generic black-box optimizer; instead, we  demonstrate the deep integration of its "strategy engine" with the gm/Id "physics-based handbook."

In this framework, the LLM is responsible for formulating macro-level optimization strategies (e.g., "The current priority is to improve the phase margin while carefully controlling power consumption"), while the gm/Id data provides quantitative, physics-based computational support for the LLM's strategies (e.g., "To implement this strategy, calculate the required W for M1 from the gm/Id table to be XX~$\mu$m").

The main contributions of this paper are as follows:
\begin{itemize}
    \item Designed an innovative synergistic reasoning framework that combines the strategic reasoning capabilities of LLMs with the physical precision of the gm/Id method.
    \item Implemented a complete closed-loop automated workflow that guides the LLM to make transparent and interpretable decisions using Chain-of-Thought (CoT).
    \item Validated the framework's effectiveness through a series of experiments, including its efficient convergence at TT and across PVT corners, and benchmarked it against traditional methods.
    \item Scientifically proved through an ablation study that the synergy of LLM's strategic reasoning and gm/Id's quantitative data is indispensable for efficient and successful automated design, achieving a "\(1+1>2\)" effect.
\end{itemize}

This paper will use the design of a classic two-stage operational amplifier as a case study to elaborate on and validate our proposed synergistic reasoning framework. We demonstrate through experiments that this framework enables an LLM to perform efficient and interpretable reasoning, progressively optimizing circuit parameters in an automated flow, and ultimately meeting all design specifications with a design quality and efficiency that is comparable to, and in some aspects superior to, that of a senior engineer.

The remainder of this paper is organized as follows:
\begin{itemize}
    \item \textbf{Section~\ref{sec:framework}} details the composition of the synergistic reasoning framework, explaining how the LLM functions as a strategy engine using CoT and how the gm/Id method serves as a physical fact-base to provide quantitative support, and elaborates on their synergistic workflow.
    \item \textbf{Section~\ref{sec:experiments}} presents comprehensive experimental validation, including the framework's performance at the typical (TT) and across all PVT corners, and scientifically demonstrates the necessity of the synergy between LLM strategic reasoning and gm/Id quantitative data through a crucial ablation study. It concludes with a benchmark comparison against manual design.
    \item \textbf{Section~\ref{sec:discussion}} provides an in-depth discussion of the experimental results, analyzes the fundamental advantages of synergistic reasoning, and explores the method's limitations and future research directions.
    \item \textbf{Section~\ref{sec:conclusion}} summarizes the entire paper.
\end{itemize}

\section{The Synergistic Reasoning Framework}
\label{sec:framework}

\subsection{Framework Overview}
Our synergistic reasoning framework is a closed-loop automated system implemented via Python scripts, which organically coordinates the interaction between a Large Language Model (LLM) API and a circuit simulator (Ngspice in this work). It aims to mimic and surpass the traditional manual design process. The core idea of the framework is to combine the macro-level strategic planning capabilities of the LLM with the micro-level physical precision of the gm/Id method, creating an automated design partner that understands both "strategy" and "tactics."

\begin{figure}[htbp]
\centerline{\includegraphics[width=\columnwidth]{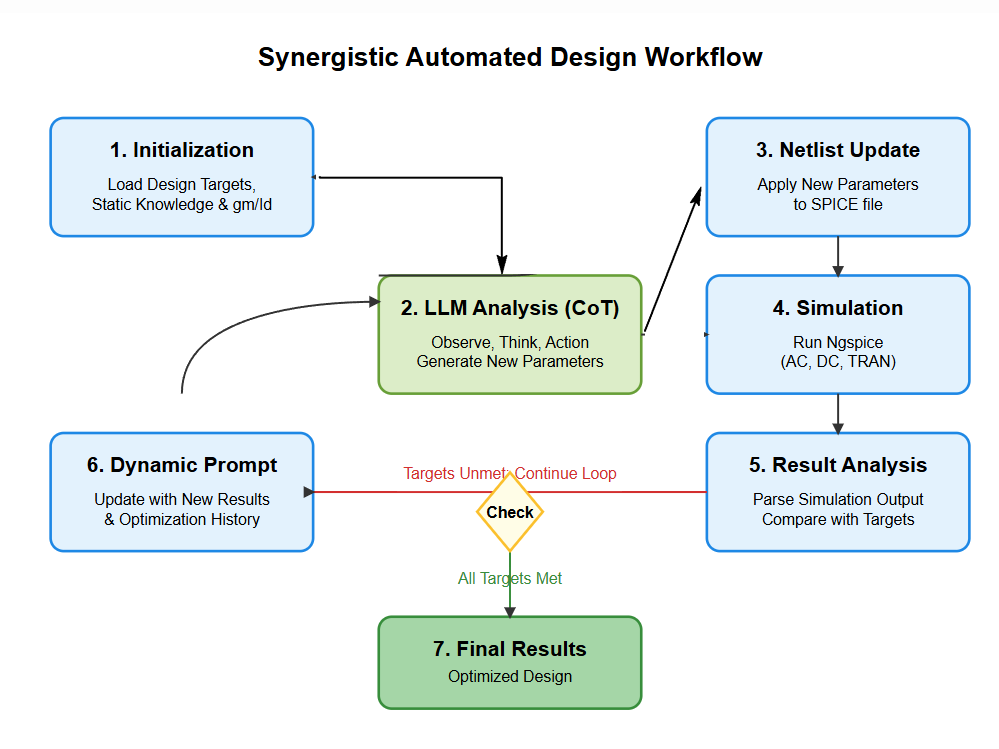}}
\caption{The Synergistic Automated Design Workflow.}
\label{fig:workflow}
\end{figure}

The entire workflow follows an iterative optimization loop, with the following steps:
\begin{enumerate}
    \item \textbf{Initialization}: The system calls the LLM with an initial prompt containing the design targets and the static knowledge base (including gm/Id data), and the LLM generates the first version of the circuit parameters.
    \item \textbf{Update \& Simulation}: The script automatically parses the circuit parameters returned by the LLM, updates the circuit netlist file, and invokes Ngspice to perform various analyses such as AC, DC, and transient.
    \item \textbf{Analysis \& Evaluation}: The script automatically extracts key performance metrics from the simulation output and compares them against the design specifications.
    \item \textbf{Dynamic Feedback \& Iterative Reasoning}:If specifications are not met, the system integrates current simulation results, circuit parameters, and optimization history into a new dynamic context for the LLM. This initiates a new \textit{reasoning-generation-simulation-evaluation} cycle until all metrics are satisfied or the iteration limit is reached.
\end{enumerate}

\subsection{The LLM as a Strategy Engine: Application of Chain-of-Thought (CoT)}
In this framework, the core role of the LLM is to act as a "strategy engine" capable of complex trade-offs. To make its decision-making process more transparent, interpretable, and logical, we guide the LLM to reason using the \textbf{Chain-of-Thought (CoT)}~\cite{wei2022chain} method.

To empower the LLM to perform effective, context-aware strategic reasoning, we designed a Dynamic Prompt engineering mechanism. This prompt consists of two parts:
\begin{itemize}
    \item \textbf{Static Knowledge Base}: Provided during the first interaction, this includes the basic operating principles of the target circuit, design heuristics, and the gm/Id lookup tables that serve as the physical fact-base. This part remains constant throughout the optimization process, forming the LLM's "long-term memory" and "knowledge base."
    \item \textbf{Dynamic Iteration Context}: This is dynamically generated and updated in each iteration. It includes the complete parameters from the previous round (Netlist), the simulation results, a summary of the optimization history (Performance History), and the design specifications that are currently unmet. This part provides the core information needed for the LLM's CoT reasoning to perform the "Observation" step, forming its "short-term memory" or "workbench."
\end{itemize}

\begin{figure}[htbp]
\centerline{\includegraphics[width=\columnwidth]{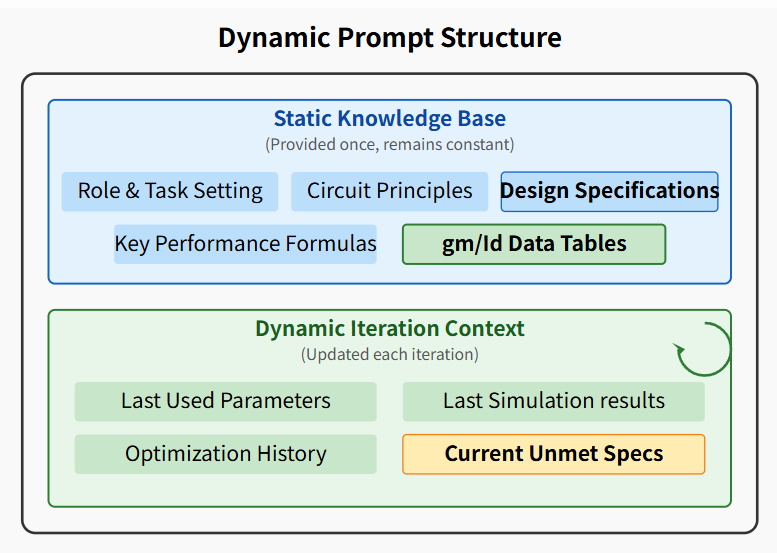}}
\caption{The structure of the Dynamic Prompt.}
\label{fig:prompt}
\end{figure}

We also introduced the "Output Visibility" principle in our prompt engineering: we explicitly require that each output from the LLM must not only provide the final netlist but also present its detailed thinking and reasoning process. We found that this requirement for process transparency encourages the LLM to conduct deeper analysis and trade-offs, thereby improving the quality of its decisions and preventing it from giving random or ill-considered answers.

Based on this dynamic input and output requirement, the LLM is guided to reason in a CoT manner. As shown in Fig.~\ref{fig:cot}, this reasoning process forms a complete "Observe-Think-Action" loop:

\begin{figure}[htbp]
\centerline{\includegraphics[width=\columnwidth]{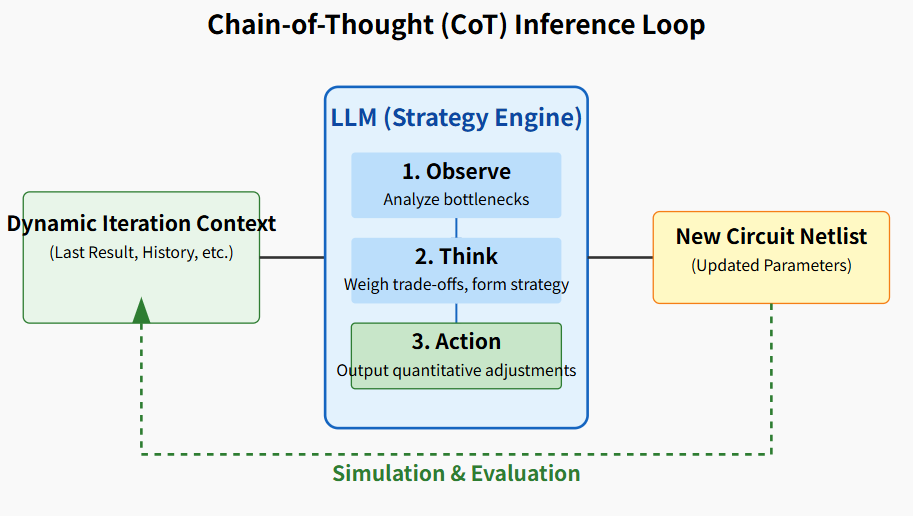}}
\caption{The Chain-of-Thought (CoT) Inference Loop.}
\label{fig:cot}
\end{figure}

\begin{enumerate}
    \item \textbf{Observation}: "What do I see?" In this step, the LLM is required to analyze the simulation results from the current round, compare them with the previous round and the design targets, and clearly identify which metrics have improved, which have worsened, and what the main performance bottleneck is at present.
    \item \textbf{Thinking Process}: "Based on my observation, what should I do?" This is the core of strategic reasoning. The LLM needs to demonstrate its ability to make trade-offs. For example, the LLM might perform the following strategic reasoning: "The Phase Margin (PM) is still too low. Although I can simply increase the compensation capacitor Cc to improve PM, this will sacrifice the Gain-Bandwidth Product (GBW). A better strategy might be to increase Cc while moderately increasing the transconductance gm7 of the output stage transistor to push the second pole to a higher frequency, thereby compensating for the loss in GBW."
    \item \textbf{Action}: "What are my specific instructions?" In this step, the LLM translates the strategy derived from its thinking process into a set of specific, executable parameter adjustment instructions.
\end{enumerate}

This CoT structure not only makes the entire optimization process interpretable and traceable but also significantly enhances the LLM's logic and success rate in multi-objective optimization.

\subsection{The gm/Id Method as a Physical Fact-Base: The Cornerstone of Quantitative Decisions}
If CoT gives the LLM the ability to "think," then the gm/Id data provides the physical precision for its "actions." In this framework, gm/Id is not just a design concept but is implemented as a "physics-based handbook" that the LLM can directly query.

In practice, we first generate a series of Lookup Tables (LUTs)~\cite{jespers2017systematic} by performing extensive SPICE sweeps on the target process library (e.g., Skywater 130nm). These tables record key performance parameters of the transistors, such as transconductance (gm), output conductance (gds), and gate-source capacitance (Cgg), as well as their core ratio gm/Id, under different bias currents (Id), channel lengths (L), and gate-source voltages (Vgs).

In each iteration, when the LLM formulates a strategy during its thinking phase, such as \textit{need to increase the gm of M1 by 15\%}, its task in the action phase shifts from vague instructions like \textit{increase the width W of M1} to precise quantitative operations. The LLM queries the provided gm/Id lookup tables to calculate the exact W value required to achieving a 15\% increase in $g_m$ at the current bias current, or selects an appropriate impedance ($1/g_{ds}$) based on gain requirements.

This mechanism "anchors" the LLM's reasoning to physical reality, ensuring that its decisions are scientific and achievable, and fundamentally eliminating the possibility of guessing.

\subsection{The Synergistic Workflow: A Detailed Look at One Iteration}
Let's walk through a typical design scenario to specifically demonstrate the synergistic workflow of the "brain" and the "handbook" within one iteration cycle:
\begin{enumerate}
    \item \textbf{LLM Receives Feedback}: Assume the LLM receives a dynamic prompt containing information such as "\(\mathrm{PM} = 55^{\circ}\) (\(\text{target} > 60^{\circ}\)), \(\mathrm{GBW} = 22\,\mathrm{MHz}\) (\(\text{target} > 20\,\mathrm{MHz},\ 10\%\ \text{margin}\)), \(\mathrm{SR} = 24\,\mathrm{V}/\mu\mathrm{s}\) (\(\text{target} > 20\,\mathrm{V}/\mu\mathrm{s},\ 20\%\ \text{margin}\))."
    
    \item \textbf{Observation}: The LLM analyzes and points out: "The most significant issue at present is the insufficient Phase Margin (PM). Meanwhile, although the Gain-Bandwidth Product (GBW) and Slew Rate (SR) meet the requirements, their design margins are limited."
    \item \textbf{Thinking Process}: This is the core of strategic reasoning. The LLM will conduct the following trade-off analysis:
    \begin{itemize}
        \item "The most direct way to improve PM is to increase the compensation capacitor Cc. However, according to the formulas `GBW $\approx$ gm1/2$\pi$Cc` and `SR $\approx$ I\_tail/Cc`, increasing Cc will simultaneously decrease both GBW and SR."
        \item "Checking the current margins: GBW has only a 10\% margin, and SR has a 20\% margin. This means I cannot increase Cc without limit, otherwise GBW and SR might fall out of specification."
        \item "Therefore, a more robust and synergistic strategy is: (a) Slightly increase Cc, using a portion of the GBW and SR margins to gain some improvement in PM; (b) The remaining PM deficit requires extending the second pole to higher frequencies. This can be most effectively accomplished by increasing the transconductance $g_{m7}$ of the output stage transistor (M7).
    \end{itemize}
    \item \textbf{Action (Synergizing with gm/Id)}: Based on the above thinking, the LLM translates this complex strategy into a set of specific, quantitative instructions:
    \begin{itemize}
        \item "`Cc`: Slightly increase the value from [1.0\,pF] to [1.1\,pF]."
        \item "`M7`: To improve `gm7`, its bias current or W/L ratio needs to be increased. Query the gm/Id table, and calculate that to increase `gm7` by [15\%] while maintaining the current Vov, the number of parallel fingers m for `M7` needs to be increased from [8] to [10], and the mirror ratio m of the current mirror `M6` needs to be adjusted accordingly."
    \end{itemize}
    \item \textbf{System Execution}: The Python script parses these instructions, updates the netlist, and proceeds to the next round of simulation.
\end{enumerate}
This example clearly shows that the LLM is not executing rigid rules like `if $PM < 60$ then $Cc = Cc * 1.1$`, but is instead performing dynamic, forward-looking resource allocation and strategic trade-offs within a multi-dimensional constraint space, demonstrating sophisticated reasoning capabilities.

\section{Experiments and Analysis}
\label{sec:experiments}

To scientifically and comprehensively validate the effectiveness and efficiency of our proposed "synergistic reasoning" framework, this chapter will present and analyze a series of experiments.

\begin{figure}[htbp]
\centerline{\includegraphics[width=0.9\columnwidth]{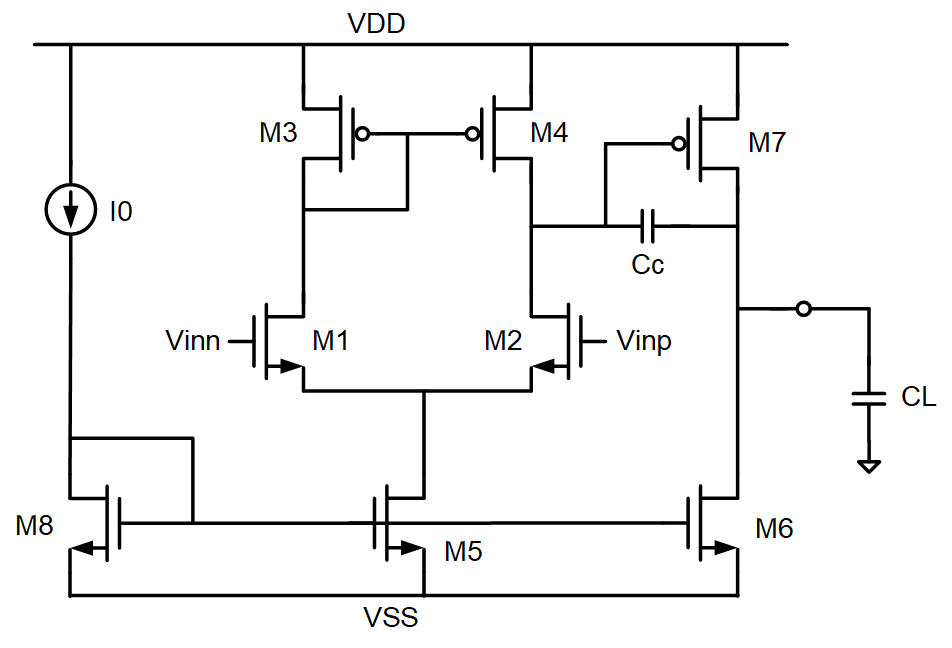}}
\caption{Schematic of the two-stage operational amplifier used as the design case.}
\label{fig:opamp_schematic}
\end{figure}

We have chosen a classic, widely studied two-stage operational amplifier as the design case for this validation. Its schematic is shown in Fig.~\ref{fig:opamp_schematic}. This choice is based on three main considerations:
\begin{itemize}
    \item \textbf{a) Core Representativeness}: The operational amplifier is one of the most fundamental and core building blocks in analog integrated circuits. Its applications are ubiquitous, and its performance directly affects numerous analog and mixed-signal systems.
    \item \textbf{b) Typical Design Challenge}: The op-amp design process is a typical multi-objective optimization problem. The designer must make complex trade-offs among multiple conflicting performance metrics, such as Gain, Gain-Bandwidth Product (GBW), Phase Margin (PM), Slew Rate (SR), and Power.
    \item \textbf{c) Universally Recognized Test Benchmark}: Precisely because of its ubiquity and design challenges, the two-stage op-amp has been recognized as a standard "litmus test" in this field. It is used as the core experimental subject in numerous academic publications on circuit automation to demonstrate the effectiveness of new methods.
\end{itemize}

\subsection{Experimental Setup and Design Targets}
All experiments in this study were conducted based on the following environment:
\begin{itemize}
    \item \textbf{LLM Model}: Google Gemini (gemini-2.0-flash)
    \item \textbf{Circuit Simulator}: Ngspice (Version 42)
    \item \textbf{Process Library}: Skywater 130nm PDK
\end{itemize}

\begin{table}[htbp]
\caption{Design Performance Specifications}
\begin{center}
\begin{tabular}{cc}
\toprule
\textbf{Performance Metric} & \textbf{Requirement} \\
\midrule
Open-loop Gain (Gain) & $>$ 60 dB \\
Gain-Bandwidth Product (GBW) & $>$ 20 MHz \\
Phase Margin (PM) & $>$ 60$^{\circ}$ \\
Slew Rate (SR) & $>$ 20 V/$\mu$s \\
Power Consumption (Idc) & $<$ 200 $\mu$A \\
\bottomrule
\multicolumn{2}{l}{Note: Load capacitance $C_L = 2$\,pF (fixed)}
\end{tabular}
\label{tab:specs}
\end{center}
\end{table}

\subsection{Experiment 1: Rapid Convergence at Typical-Typical (TT) Corner}
In this experiment, we fed the initial prompt into the framework to start the automated design process. The entire process aims to demonstrate the framework's capability to achieve convergence from scratch at the typical (TT) corner.

\subsubsection{Initial Design (Iteration 1)}
Unlike "black box" algorithms, our framework does not start with a random guess. The LLM first acts as a "theoretical calculator," applying circuit design formulas and using the provided gm/Id data to make detailed, physics-based parameter estimations. This ensures the initial parameters are reasonable. However, the first simulation showed that Gain (52.19 dB), GBW (17.58 MHz), and PM (58.49 deg) all failed to meet the targets, kicking off the iterative optimization process.

\subsubsection{Iterative Optimization: Non-linear Strategic Trade-offs}
The subsequent iterations fully demonstrate the LLM's ability to perform complex, non-linear trade-offs:
\begin{itemize}
    \item \textbf{Iteration 2}: Facing multiple failures, the LLM formulated a multi-pronged synergistic strategy, adjusting C1, M1/M2, and M7 simultaneously to address different metrics.
    \item \textbf{Iteration 3}: Observing that GBW (43.29 MHz) was far exceeding the target while PM (54.14$^{\circ}$) was still insufficient, the LLM demonstrated its trade-off capability. It made an opposite move to the previous step, significantly increasing C1 and decreasing M1/M2's width to actively reduce GBW and prioritize stability.
    \item \textbf{Iteration 4}: With only Gain (48.41 dB) remaining as the last issue, the LLM's strategy refocused, taking the most direct and effective approach by increasing the channel length of key transistors and the multiplier of the output stage.
    \item \textbf{Iteration 5}: At this point, with only GBW (19.16 MHz) slightly below target, the LLM performed a final fine-tuning, making a minimal adjustment to C1 and M1/M2's width to precisely push GBW over the threshold without compromising other metrics.
\end{itemize}
After 5 iterations, all performance metrics were met at the TT corner.

\begin{table}[htbp]
\caption{Final Optimization Results at TT Corner}
\begin{center}
\begin{tabular}{cccc}
\toprule
\textbf{Metric} & \textbf{Target} & \textbf{Final Result (TT)} & \textbf{Status} \\
\midrule
Gain & $>$ 60.0 dB & 62.40 dB & \checkmark \\
GBW & $>$ 20.0 MHz & 25.30 MHz & \checkmark \\
PM & $>$ 60.0$^{\circ}$ & 65.80$^{\circ}$ & \checkmark \\
SR & $>$ 20.0 V/$\mu$s & 28.18 V/$\mu$s & \checkmark \\
Idc & $<$ 200.0 $\mu$A & 116.4 $\mu$A & \checkmark \\
\bottomrule
\end{tabular}
\label{tab:tt_results}
\end{center}
\end{table}

\begin{table}[htbp]
\caption{Key Parameter Changes During Optimization Iterations}
\begin{center}
\resizebox{\columnwidth}{!}{%
\begin{tabular}{lcccccc}
\toprule
\textbf{Param.} & \textbf{Iter 1} & \textbf{Iter 2} & \textbf{Iter 3} & \textbf{Iter 4} & \textbf{Iter 5 (TT)} & \textbf{Iter 6 (All-Corner)} \\
\midrule
C1 & 1.0p & 0.7p & 1.2p & 1.2p & 1.0p & \textbf{0.9p} \\
M1/M2 (L/W) & 0.18/1 & 0.18/1 & 0.18/3 & 0.5/3 & 0.5/4 & \textbf{0.5/4} \\
M3/M4 (L/W) & 0.18/1 & 0.18/1 & 0.18/1 & 0.5/1 & 0.5/1 & \textbf{0.5/1} \\
M5 (m) & 2 & 3 & 3 & 3 & 3 & \textbf{3} \\
M7 (m) & 4 & 6 & 6 & 8 & 8 & \textbf{8} \\
M6 (m) & 8 & 8 & 8 & 8 & 8 & \textbf{8} \\
\bottomrule
\multicolumn{7}{l}{Note: L/W units are in $\mu$m.}
\end{tabular}
}
\label{tab:param_changes}
\end{center}
\end{table}

The entire process clearly shows that the LLM's reasoning is not simple linear or if-else logic, but a dynamic, multi-dimensional feedback-based process of constantly adjusting optimization strategies. To demonstrate this more intuitively, we excerpt the CoT log from the 3rd iteration in Fig.~\ref{fig:cot_log}.

\begin{figure}[htbp]
\begin{lstlisting}[
  frame=single,
  basicstyle=\ttfamily\small,
  breaklines=true,
  label={fig:cot_log}
]
**********************************************
 LLM Chain-of-Thought (CoT) Log - Iteration 3
**********************************************
[Context]: 
LLM receives results from iter 2: 
Gain=51.16dB(fail), GBW=43.29MHz(>>target), 
PM=54.14deg(fail), SR=40.47V/us(pass).

[LLM Output]:
1. Observation:
   "Current issues: Gain is insufficient, Phase 
   Margin is insufficient. GBW is too high, 
   but SR is met."

2. Thinking Process:
   "The insufficient Phase Margin indicates that 
   the poles/zeros are too close to the GBW. 
   Since GBW is much larger than the target, 
   reducing GBW is an effective way to improve 
   Phase Margin..."

3. Action:
   (Formulates a strategy to trade "excess" 
   GBW for "PM")
   "C1: Increase C1 from 0.7p to 1.2p..."
   "M1, M2: Decrease the W of M1 and M2 to 
   lower gm1, further reducing GBW..."

\end{lstlisting}
\caption{An example of the CoT log from the 3rd iteration, showing strategic trade-offs.}
\label{fig:cot_log}
\end{figure}

\subsection{Experiment 2: Robust Optimization Across PVT Corners}
After the design successfully converged at the TT corner, our framework automatically proceeded to the more stringent all-corner (PVT) validation. A  core specification factors system sets slightly relaxed, more physically realistic design targets for different process corners.

After the 5th iteration, the all-corner simulation revealed that the GBW of the SF corner (17.92 MHz) failed to meet its 19.0 MHz target. Faced with this specific failure, the framework did not start over but fed the detailed failure information back to the LLM.

\begin{itemize}
    \item \textbf{Iteration 6}: The LLM observed that the TT corner had ample GBW margin (25.3 MHz) while only the SF corner's GBW was slightly below spec. It reasoned that the safest strategy was a micro-adjustment with a global positive effect. It made an extremely concise decision—to only fine-tune C1 from 1.0\,pF to 0.9\,pF, keeping all other parameters unchanged.
\end{itemize}
After this micro-adjustment, another all-corner simulation showed that all performance metrics under all process corners successfully met their respective requirements.

\begin{table}[htbp]
\caption{Worst-Case Performance Across All PVT Corners}
\begin{center}
\begin{tabular}{ccccc}
\toprule
\textbf{Metric} & \textbf{Worst-Case Target} & \textbf{Result} & \textbf{Corner} & \textbf{Status} \\
\midrule
Gain & $>$ 54.0 dB & 59.90 dB & FS & \checkmark \\
GBW & $>$ 19.0 MHz & 19.79 MHz & SF & \checkmark \\
PM & $>$ 54.0$^{\circ}$ & 63.00$^{\circ}$ & FF & \checkmark \\
SR & $>$ 18.0 V/$\mu$s & 30.59 V/$\mu$s & FS & \checkmark \\
Idc & $<$ 240.0 $\mu$A & 118.3 $\mu$A & SF & \checkmark \\
\bottomrule
\end{tabular}
\label{tab:pvt_results}
\end{center}
\end{table}

\begin{table*}[htbp]
\caption{Ablation Study Results Comparison}
\begin{center}
% Using p{width} for columns with long text to enforce line wrapping.
% This method does not require any extra packages.
\begin{tabular}{p{2.8cm} p{2.0cm} p{2.5cm} c c p{6.5cm}}
\toprule
\textbf{Experiment Group} & \textbf{Converged (All Corners)} & \textbf{Iterations to Converge} & \textbf{FOM (TT)} & \textbf{FoMA (TT)} & \textbf{Typical Failure Mode Analysis} \\
\midrule
\textbf{Group A (This Work)} & Yes & 6 (TT:5, PVT:1) & \textbf{265.7} & \textbf{8.46} & None, converged efficiently and directly. \\
\addlinespace % Adds a little extra vertical space for readability
Group B (LLM w/o gm/id data) & No & 16 (TT:15, PVT:failed) & 119.7 & 0.73 & Lacks quantitative guidance, improper adjustment magnitudes lead to performance oscillation and extremely slow convergence. \\
\addlinespace
Group C (script w/i gm/id data) & No & N/A (stuck in a loop) & N/A & N/A & Rigid rules cannot handle complex multi-objective trade-offs, easily gets trapped in local optima or parameter oscillation. \\
\bottomrule
\end{tabular}
\label{tab:ablation}
\end{center}
\end{table*}

\subsection{Experiment 3: Ablation Study - Validating the Necessity of Synergy}
To scientifically validate the necessity of the synergy between the LLM's strategic reasoning (the "brain") and the gm/Id quantitative data (the "handbook"), we designed three sets of ablation experiments, as detailed in Table~\ref{tab:ablation}.

\textbf{Group B (LLM without gm/Id data)} performed drastically differently from our main approach. Without precise gm/Id data, the LLM's initial design had a phase margin of only 13.47 degrees. In subsequent iterations, although it understood the qualitative direction of adjustments, its parameter changes became blind, resulted in inappropriate parameter adjustments. It took 15 iterations to barely converge at the typical-typical (TT) corner and ultimately failed the all-corner validation.

\textbf{Group C (gm/Id data without LLM)}, a rule-based script, also failed to converge. Its rigid rules like `if $PM < 60$: Cc = Cc * 1.1` could not handle complex trade-offs and often fixed one metric by breaking another, getting stuck in a parameter adjustment "deadlock".

The results powerfully demonstrate that the LLM's "brain" and the gm/Id "handbook" are indispensable. Only by synergizing the two (Group A) can truly efficient and intelligent automated design be achieved.

\subsection{Experiment 4: Framework Universality Verification}
To verify that our framework's success is not dependent on a single LLM, we repeated the experiment with two other mainstream models: \textbf{DeepSeek (v3)} and \textbf{Qwen (qwen-max)}.

The results were encouraging. As shown in Table~\ref{tab:llm_compare}, although the models showed differences in reasoning style (e.g., Qwen's aggressiveness vs. Gemini/DeepSeek's stability), they all ultimately succeeded in driving the design to convergence, meeting all specifications including process corners. This experiment demonstrates the powerful generalization ability of our framework itself: it provides a universal "socket" that can effectively utilize the reasoning capabilities of various advanced LLMs.

\begin{table}[htbp]
\caption{Performance Comparison of Different LLM Models}
\begin{center}
\begin{tabular}{lcccc}
\toprule
\textbf{LLM Model} & \textbf{Total Iters.} & \textbf{FOM (TT)} & \textbf{FoMA (TT)} \\
\midrule
\textbf{Gemini-2.0-flash} & \textbf{6} & \textbf{265.7} & \textbf{8.46} \\
DeepSeek-V3 & 10 & 159.5 & 0.82 \\
Qwen-Max & 9 & 129.9 & 0.79 \\
\bottomrule
\end{tabular}
\label{tab:llm_compare}
\end{center}
\end{table}

\subsection{Experiment 5: Performance and Efficiency Benchmarking}
To measure the practical value of our framework, we benchmarked its automated design results against the manual design of a senior engineer.

\begin{table}[htbp]
\caption{Comparison Between Automated and Manual Design}
\begin{center}
\begin{tabular}{lcc}
\toprule
\textbf{Comparison Item} & \textbf{This Work (Gemini)} & \textbf{Senior Engineer} \\
\midrule
Total Design Time & $\sim$5 minutes & Several hours (typical) \\
Final FOM & \textbf{265.7} & 226.7 \\
Final FoMA & \textbf{8.46} & 5.30 \\
Final Area (est.) & \textbf{31.4 $\mu$m$^2$} & 42.8 $\mu$m$^2$ \\
\bottomrule
\end{tabular}
\label{tab:human_compare}
\end{center}
\end{table}

The results in Table~\ref{tab:human_compare} show advantages on two levels. First, in efficiency, the automated framework reduces the design cycle from several hours to \textbf{about 5 minutes}. Second, in design quality, the framework's solution is 27\% smaller in area and has 17\% and 60\% higher FOM and FoMA metrics, respectively.

This superiority in PPA (Power, Performance, Area) stems from a difference in design philosophy. Engineers often adopt more conservative, larger-area designs out of consideration for second-order effects like device mismatch, a reasonable engineering decision. In contrast, our framework, due to its efficient and targeted iteration capability, has the ability to continue exploring after meeting basic specs to find a more PPA-optimal solution—a process often infeasible for manual design due to its prohibitive time cost.

\section{Discussion}
\label{sec:discussion}

\subsection{The Nature and Advantage of Synergistic Reasoning}
The core of this research is the validation of a "synergistic reasoning" model. The experimental results, especially the ablation study, clearly reveal the advantages of this model. Traditional automation methods, whether they are pure search algorithms or pure qualitative reasoning, have their inherent flaws. Our framework, by combining the flexible strategic planning capabilities of the LLM with the physical precision of the gm/Id method, creates an automated designer that understands both "strategy" and "tactics," thereby achieving a breakthrough in design efficiency and quality.

\subsection{From "Black-Box Optimization" to "White-Box Reasoning": A Paradigm Shift}
The deeper value of this research lies not just in proposing a more efficient automation tool, but in exploring and realizing a paradigm shift from "black-box optimization" to \textbf{"white-box reasoning"}.

The optimization process of traditional automation algorithms is an opaque "black box" to the designer, making it difficult to build trust. Our synergistic reasoning framework successfully opens this "black box" through two core mechanisms, making it an interpretable and verifiable "white box."

The first mechanism is interpretability, provided by Chain-of-Thought (CoT). As shown in Fig.~\ref{fig:cot_log}, the LLM details its entire \textit{observe-think-action} process. The designer can clearly see why the LLM made a certain decision. This process transparency makes the AI's reasoning understandable and analyzable, providing unprecedented insight into automated decision-making.

The second mechanism is verifiability, enabled by the gm/Id methodology. The LLM's reasoning process provides interpretability at the macro-strategy level, while its micro-level quantitative decisions remain traceable and verifiable. When the LLM determines transistor transconductance adjustments, these decisions are grounded in precise calculations using the provided gm/Id lookup tables rather than heuristic approximations.

In summary, the combination of an "interpretable" strategy engine (LLM + CoT) and a "verifiable" physics-based handbook (gm/Id) jointly builds this new paradigm of "white-box reasoning." This is crucial for the application of AI in serious, high-reliability engineering design fields, and it marks a solid step towards a truly trustworthy and collaborative "AI design partner."

\subsection{Limitations and Future Work}
Although this framework has successfully validated the feasibility of synergistic reasoning, some limitations still exist, which also point to directions for future research:
\begin{enumerate}
    \item \textbf{Topology Limitation}: The current work focuses on parameter optimization of a fixed circuit topology. Enabling the LLM to assist in circuit topology selection, modification, and innovation represents a more challenging and valuable research direction.
    
    \item \textbf{Physical Effect Modeling}: As discussed in Section 3.6, engineers consider second-order physical effects such as device mismatch and flicker noise. Our current physics-based knowledge repository has not yet incorporated these effects. Future work could explore integrating more comprehensive physical models into the framework.
    
    \item \textbf{Application Scope}: This paper conducted in-depth validation using a two-stage operational amplifier. The next step involves extending this framework to more complex analog modules, including low-dropout regulators (LDOs), bandgap references, and phase-locked loops (PLLs).
\end{enumerate}

\section{Conclusion}
\label{sec:conclusion}
This paper addresses the challenges of low automation and inefficient black-box optimization algorithms in analog circuit design by proposing and successfully validating an innovative synergistic reasoning framework. This framework creatively integrates the macro-level strategic reasoning capabilities of Large Language Models (LLMs) with the micro-level physical precision of the gm/Id design methodology, effectively bridging the long-standing strategy and data gaps in automated design.

Through a classic two-stage operational amplifier design case study, we demonstrated the excellent performance of this framework. Experiments proved that the framework efficiently completes design tasks meeting both TT and all PVT corner specifications within a small number of iterations, while achieving PPA (Power, Performance, Area) metrics superior to typical manual engineering designs. The crucial ablation study and cross-LLM model validation scientifically demonstrated that the synergy between LLM strategic reasoning and gm/Id quantitative data constitutes the cornerstone of the framework's success, confirming the method's broad applicability.

More importantly, this research explores and realizes a paradigm shift from black-box optimization to \textit{white-box reasoning}. Through the combination of Chain-of-Thought and gm/Id data, the AI's decision-making process becomes interpretable and verifiable, paving a solid path toward building truly trustworthy AI design partners. We believe that this approach of combining general artificial intelligence with domain-specific scientific methodologies will provide sustained momentum and broad prospects for solving complex engineering design problems.

\bibliographystyle{IEEEtran}
\bibliography{references}

\begin{thebibliography}{1}
\providecommand{\url}[1]{#1}
\csname url@samestyle\endcsname
\providecommand{\newblock}{\relax}
\providecommand{\bibinfo}[2]{#2}
\providecommand{\BIBentrySTDinterwordspacing}{\spaceskip=0pt\relax}
\providecommand{\BIBentryALTinterwordstretchfactor}{4}
\providecommand{\BIBentryALTinterwordspacing}{\spaceskip=\fontdimen2\font plus
\BIBentryALTinterwordstretchfactor\fontdimen3\font minus \fontdimen4\font\relax}
\providecommand{\BIBforeignlanguage}[2]{{%
\expandafter\ifx\csname l@#1\endcsname\relax
\typeout{** WARNING: IEEEtran.bst: No hyphenation pattern has been}%
\typeout{** loaded for the language `#1'. Using the pattern for}%
\typeout{** the default language instead.}%
\else
\language=\csname l@#1\endcsname
\fi
#2}}
\providecommand{\BIBdecl}{\relax}
\BIBdecl

\bibitem{dastidar2010automated}
T.~R. Dastidar, P.~P. Chakrabarti, and P.~Ray, ``Automated analog circuit synthesis with genetic algorithms and particle swarm optimization,'' \emph{IEEE Transactions on Evolutionary Computation (T-EC)}, vol.~14, no.~4, pp. 594--613, 2010.

\bibitem{hakhamaneshi2019efficient}
K.~Hakhamaneshi, N.~Werblun, P.~Abbeel, and V.~Stojanovic, ``An efficient bayesian optimization approach for automated optimization of analog circuits,'' \emph{IEEE Transactions on Computer-Aided Design of Integrated Circuits and Systems (T-CAD)}, vol.~38, no.~11, pp. 2067--2080, 2019.

\bibitem{silveira1996gm}
F.~Silveira, D.~Flandre, and P.~G.~A. Jespers, ``A gm/id based methodology for the design of cmos analog circuits and its application to the synthesis of a silicon-on-insulator micropower ota,'' \emph{IEEE Journal of Solid-State Circuits (JSSC)}, vol.~31, no.~9, pp. 1314--1319, 1996.

\bibitem{liu2023analogxpert}
Y.~Liu \emph{et~al.}, ``Analogxpert: Automating analog topology synthesis by incorporating circuit design expertise into large language models,'' \emph{arXiv preprint arXiv:2312.11824}, 2023.

\bibitem{xu2024analogcoder}
H.~Xu \emph{et~al.}, ``Analogcoder: Analog circuit design via training-free code generation,'' \emph{arXiv preprint arXiv:2405.14918}, 2024.

\bibitem{yin2024ado}
Y.~Yin \emph{et~al.}, ``Ado-llm: Analog design bayesian optimization with in-context learning of large language models,'' \emph{arXiv preprint arXiv:2406.18770}, 2024.

\bibitem{wei2022chain}
J.~Wei \emph{et~al.}, ``Chain-of-thought prompting elicits reasoning in large language models,'' in \emph{Advances in neural information processing systems}, vol.~35, 2022, pp. 24\,824--24\,837.

\bibitem{jespers2017systematic}
P.~G.~A. Jespers and B.~Murmann, \emph{Systematic Design of Analog CMOS Circuits: Using Pre-Computed Lookup Tables}.\hskip 1em plus 0.5em minus 0.4em\relax Cambridge University Press, 2017.

\end{thebibliography}

\end{document}